\documentclass[twocolumn]{aastex62}
\usepackage{amsfonts,amsmath,amssymb,amsthm,epsfig,graphicx,float,tabularx}

\begin{document}
\title{Extreme mass loss in low-mass type Ib/c supernova progenitors}

\correspondingauthor{Samantha Wu}
\email{scwu@astro.caltech.edu }

\author{Samantha Wu}
\affiliation{California Institute of Technology, Astronomy Department, Pasadena, CA 91125, USA}

\author{Jim Fuller}
\affiliation{TAPIR, Walter Burke Institute for Theoretical Physics, Mailcode 350-17, California Institute of Technology, Pasadena, CA 91125, USA}

\begin{abstract}
   Many core collapse supernovae (SNe) with hydrogen-poor and low-mass ejecta, such as ultra-stripped SNe and type Ibn SNe, are observed to interact with dense circumstellar material (CSM). These events likely arise from the core-collapse of helium stars which have been heavily stripped by a binary companion and ejected significant mass during the last weeks to years of their lives.
   In helium star models run to days before core-collapse, we identify a range of helium core masses $\approx 2.5 \text{--}3\, M_{\odot}$ whose envelopes expand substantially due to helium shell burning while the core undergoes neon and oxygen burning. When modeled in binary systems, the rapid expansion of these helium stars induces extremely high rates of late-stage mass transfer ($\dot{M} \! \gtrsim \! 10^{-2} \, M_\odot/{\rm yr}$) beginning weeks to decades before core-collapse. We consider two scenarios for producing CSM in these systems: either mass transfer remains stable and mass loss is driven from the system in the vicinity of the accreting companion, or mass transfer becomes unstable and causes a common envelope event (CEE) through which the helium envelope is unbound. The ensuing CSM properties are consistent with the CSM masses ($\sim \! 10^{-2}-1 \, M_\odot$) and radii ($\sim \! 10^{13}-10^{16} \, {\rm cm}$) inferred for ultra-stripped SNe and several type Ibn SNe. Furthermore, systems that undergo a CEE could produce short-period NS binaries that merge in less than 100 Myr.
\end{abstract}

\section{Introduction}

Many types of core-collapse supernovae (SNe) show signs of interaction with dense circumstellar material (CSM), likely created by extreme mass loss at the end of the SN progenitor's life. Type Ibn SNe are characterized by interaction with hydrogen-poor and helium-rich CSM, which produces spectra dominated by narrow helium (He) lines and powers early-time light curves that often rise and decay quickly. 
Typical rise times of $\lesssim 15$ days and peak magnitudes of $M_R \! \sim \! -19$ to $-20$ mag in these events indicate ejecta masses $M_{\rm ej} = 1\text{--}5\, M_{\odot}$ and $^{56}$Ni masses $M_{\rm Ni} \lesssim 0.1\, M_{\odot}$ \citep[][]{Gangopadhyay2022,maeda2022,ho2021}. 
These SNe are thought to originate from massive stars that have previously lost their hydrogen envelopes, then expelled helium-rich CSM just before core-collapse.

A few events have been discovered with even lower $M_{\rm ej}$ and $M_{\rm Ni}$, classified as ultra-stripped SNe (USSNe). For example, the short decline time of type Ic SN iPTF 14gqr indicates a small ejecta mass of~$M_{\rm ej} \sim 0.2\, M_{\odot}$ \citep{De2018}, and consequently a low pre-collapse mass of~$M_{\rm He} \sim 1.6\, M_{\odot}$ (assuming a baryonic NS mass of~$M_{\rm NS} = 1.4\, M_{\odot}$). The type Ib SN 2019dge has~$M_{\rm ej} = 0.4\, M_{\odot}$, implying pre-collapse mass~$M_{\rm He} \sim 1.8\, M_{\odot}$ \citep{Yao2020}. Bright, rapidly rising early-time light curves and flash-ionized He emission in early spectra indicate extended CSM in SN iPTF 14gqr and SN 2019dge. 
Another type of interacting SNe, type Icn SNe, exhibit narrow emission lines from recombination of ionized carbon and oxygen instead of He. With comparable peak luminosities to type Ibn SNe but low $M_{\rm Ni}$ and $M_{\rm ej}$, type Icn SNe have been proposed to arise from similar channels to USSNe \citep{pellegrino2022b}.

Highly-stripped helium stars are the probable progenitors of USSNe. Stars that have lost their hydrogen envelopes after hydrogen burning through case B mass transfer form stripped stars from their He cores. Stripped stars with $M_{\rm He} \! \lesssim \! 4\, M_{\odot}$ expand again and initiate so-called case BB mass transfer in systems with final separations of less than a few $100\, R_{\odot}$ after case B mass transfer \citep[][]{habets1986,habets1986b}, thereby losing a significant amount of their He envelope as well. At core-collapse, their low pre-collapse masses can explain the requisite $M_{\rm ej}$ of USSNe \citep{Tauris2013,Tauris2015}. When the USSNe is formed from the initially less massive star in the binary, such systems are likely the most common progenitors of compact NS binaries \citep[][]{dewi2003,Tauris2013,Tauris2015}.

Previous work \citep{Tauris2013,Tauris2015,Yoon2010,Zapartas2017,Laplace2020} has modeled case BB mass transfer (MT) in detail to make predictions for mass loss and the final fate of the progenitor. Thus far, most stellar models do not predict large amounts of CSM near the progenitor system as detected in several USSNe and type Ibn SNe. 
Yet the vast majority of stripped progenitor models omit the evolution onward from oxygen/neon (O/Ne) burning, and they miss crucial physics that transpires during these final years of the star's lifetime that may explain such SN observations. 
We find that helium stars of masses $\approx \! 2.5-3 \, M_\odot$ rapidly re-expand while the core burns O/Ne, which initiates high rates of late-stage MT weeks to decades before core-collapse that may produce CSM. 


\begin{figure*}
    \includegraphics[width=0.501\textwidth]{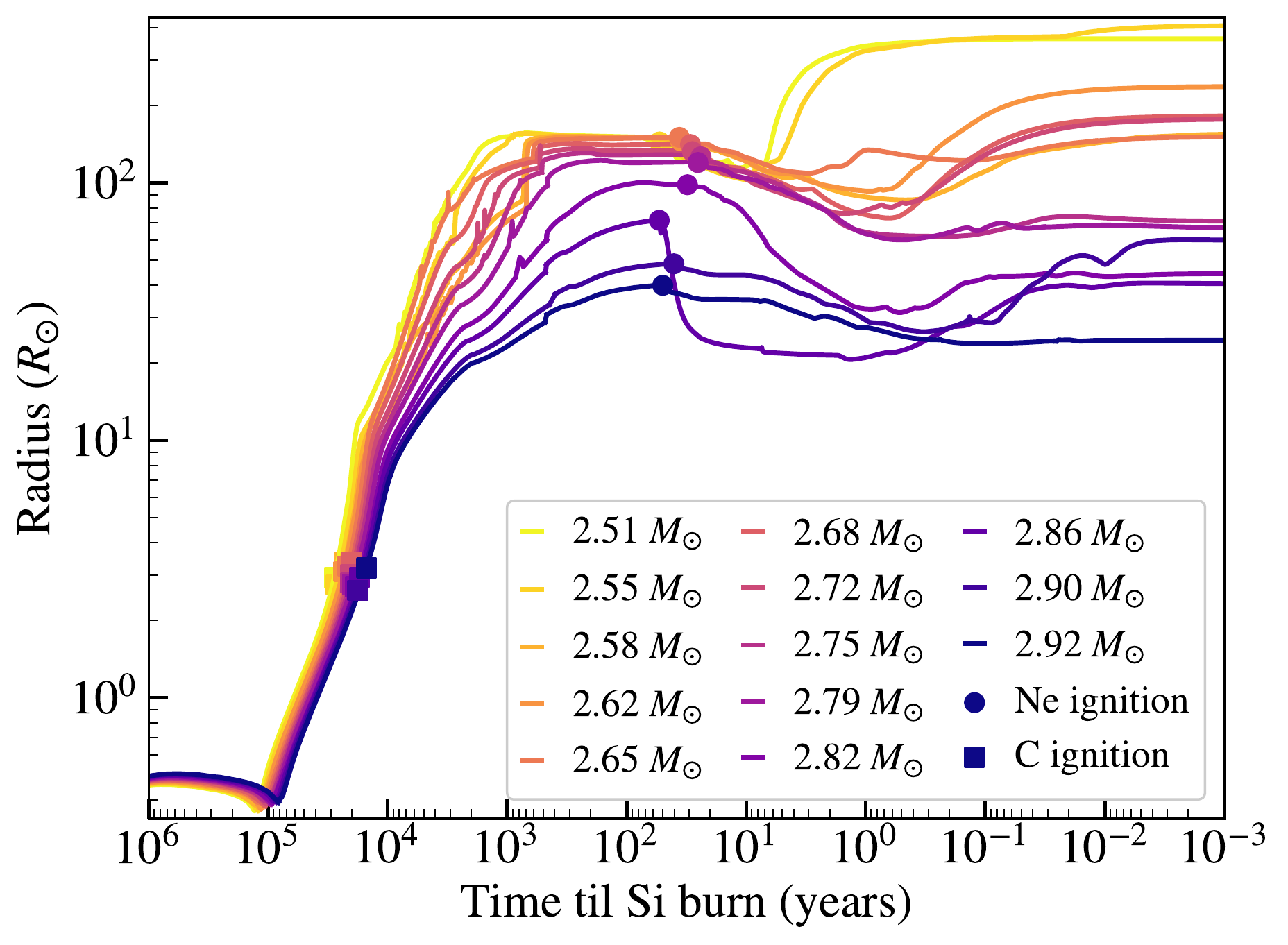}
    \includegraphics[width=0.499\textwidth]{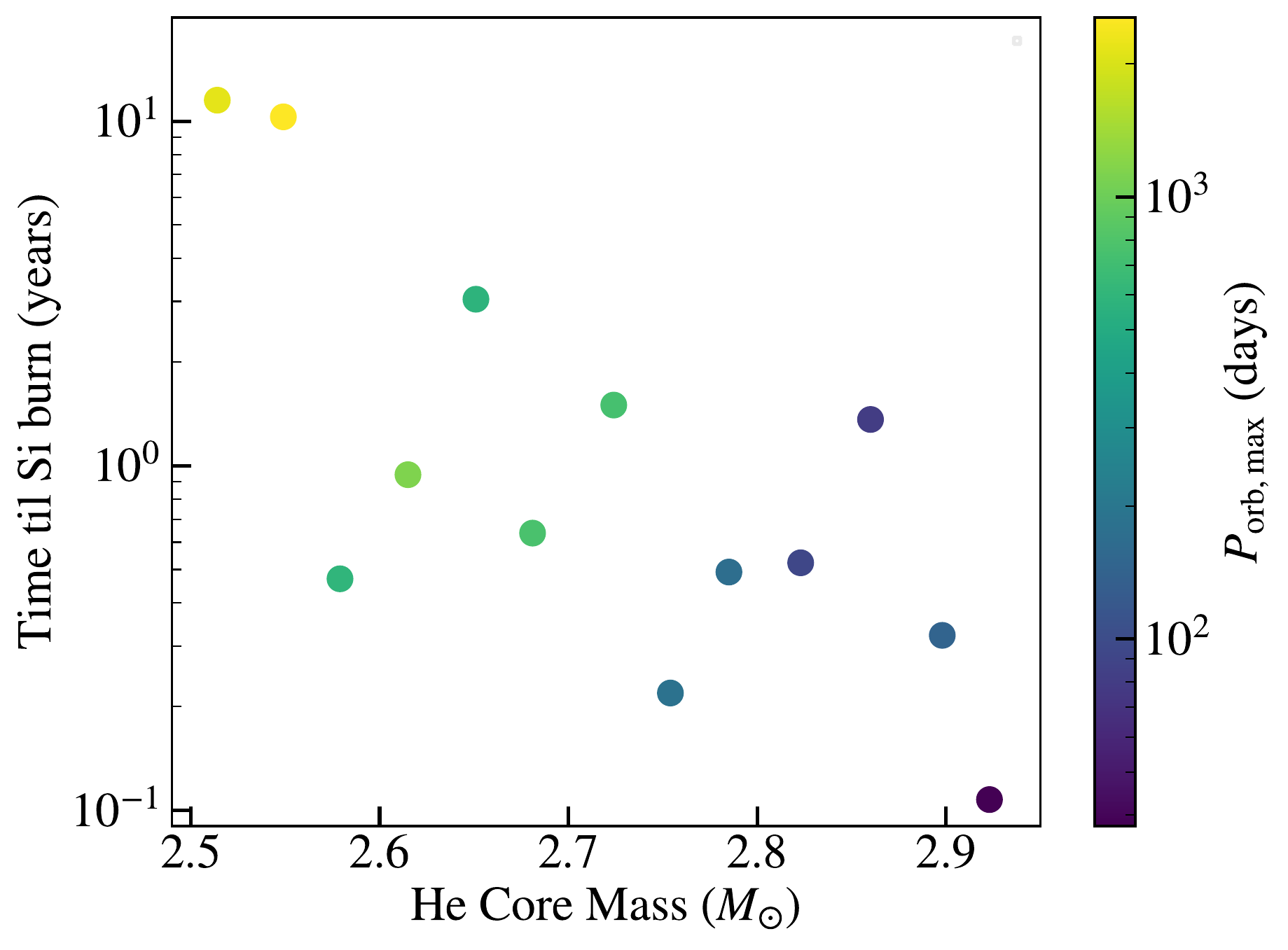}
    \caption{ Left: Evolution of the radius of single stripped stars as a function of time until Si-burning. The legend labels the initial He core mass of each stripped star. Each star expands after He burning and throughout C burning, then contracts and re-expands during O/Ne-burning. 
    Right: Time before Si-burning of the stripped stars' second expansion. Points are colored by the maximum orbital period at which the star will fill its Roche lobe in a binary with a~$1.4\, M_{\odot}$ companion star.
    }
    \label{fig:singlervst}
\end{figure*}

\section{Methods}

We use MESA \citep[version r15140,][]{mesa2011,mesa2013,mesa2015,mesa2018,mesa2019} to model 1d stellar evolution up to silicon (Si) burning of single, stripped stars at $Z = 0.02$ with $2.5\, M_{\odot} \! \lesssim \! M_{\rm He} \lesssim 3\, M_{\odot}$ \footnote{The data is available on Zenodo under an open-source 
Creative Commons Attribution license: \dataset[doi:10.5281/zenodo.7106182]{https://doi.org/10.5281/zenodo.7106182}}.
The timing of removing surface hydrogen, the amount of hydrogen remaining, and the inclusion of stellar winds each affect how initial mass $M_{\rm ZAMS}$ maps to He core mass $M_{\rm He}$ after core He burning. In this work, we remove the entire hydrogen envelope of stars with initial masses~$M_{\rm ZAMS} = 13.8 \text{--} 15\, M_{\odot}~$ once core hydrogen burning ends and evolve without wind mass loss. Ultimately, the relation between initial and helium star mass is not central to the result, and we find that the behavior of our models depends primarily on the He core mass after core He burning.
Throughout, we label our models by these initial He core masses $M_{\rm He}$. 

We also model the stripped stars in binaries at a range of orbital periods from~$1$ to~$100$ days to estimate the ensuing mass transfer rates. 
For our binary models, we consider the fiducial scenario where the stripped star formed from the initially less massive star in the binary, so its companion is a~$M_{\rm c} = 1.4\, M_{\odot}$ neutron star, represented by a point mass.
We use a modified version of the implicit MT scheme of \citet{kolb1990} for Roche lobe overflow. Since this prescription assumes an ideal gas EOS, it underestimates mass loss rates for surface layers dominated by radiation pressure; to address this, we revise the scheme to compute the pressure from the stellar model \citep[e.g.,][]{marchant2021}. We assume non-conservative mass transfer where the mass is removed from the system in the vicinity of the accretor as a fast wind. 
As we find that mass transfer rates during both case BB and late-stage mass loss are many orders of magnitude larger than the Eddington accretion limit of a NS ($\dot{M}_{\rm edd} \sim 4\times 10^{-8}\, M_{\odot}\, \rm{yr}^{-1}$), we expect that nearly~$100\%$ of the mass is lost from the system (as in, e.g., \citealt{Tauris2015}), though mass loss out the L2 and/or L3 points could modify the binary's angular momentum loss (see Section \ref{sec:discussion}).

\begin{figure*}
    \includegraphics[width=\textwidth]{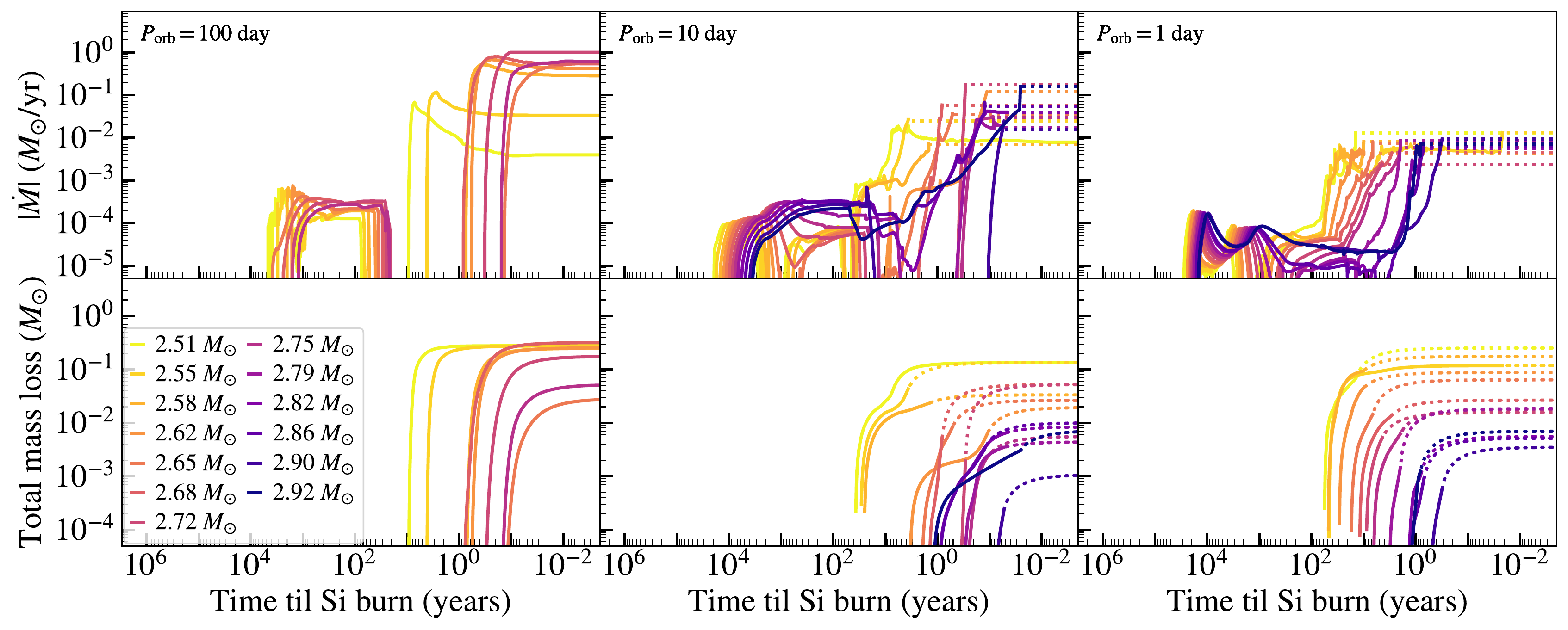}
    \caption{The mass loss rates and accumulated mass loss of the helium star models, each of which is placed in a binary with a $1.4 \, M_\odot$ compact companion at the initial orbital periods listed in the top panels. The legend indicates the initial mass of each helium star. For simulations that end before Si-burning, we assume that~$\dot{M}$ remains steady until Si-burning and extrapolate the accumulated mass loss until Si-burning, shown as dotted lines.
    }
    \label{fig:mdots}
\end{figure*}

\begin{figure*}
    \includegraphics[width=0.5\textwidth]{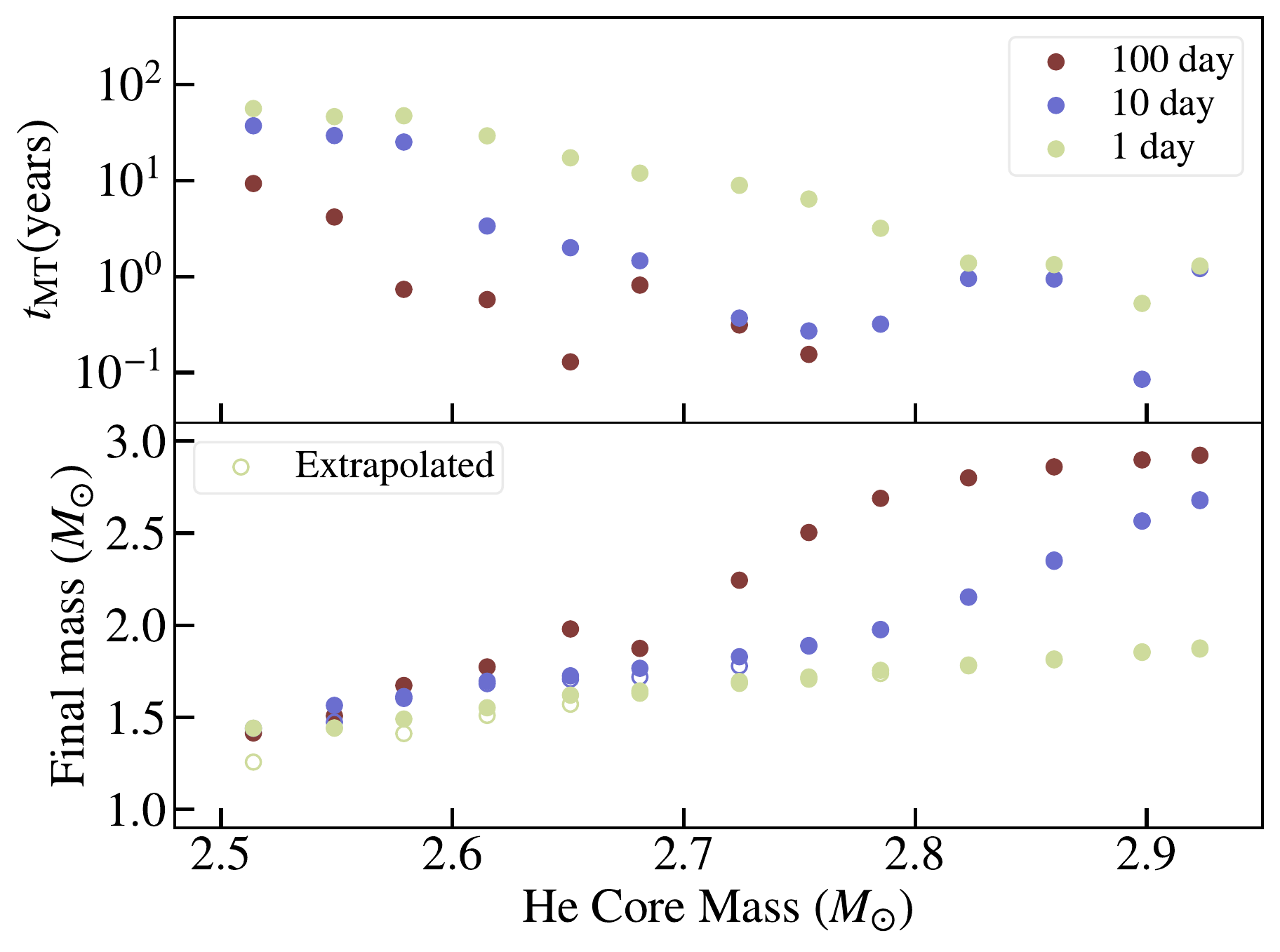}
    \includegraphics[width=0.5\textwidth]{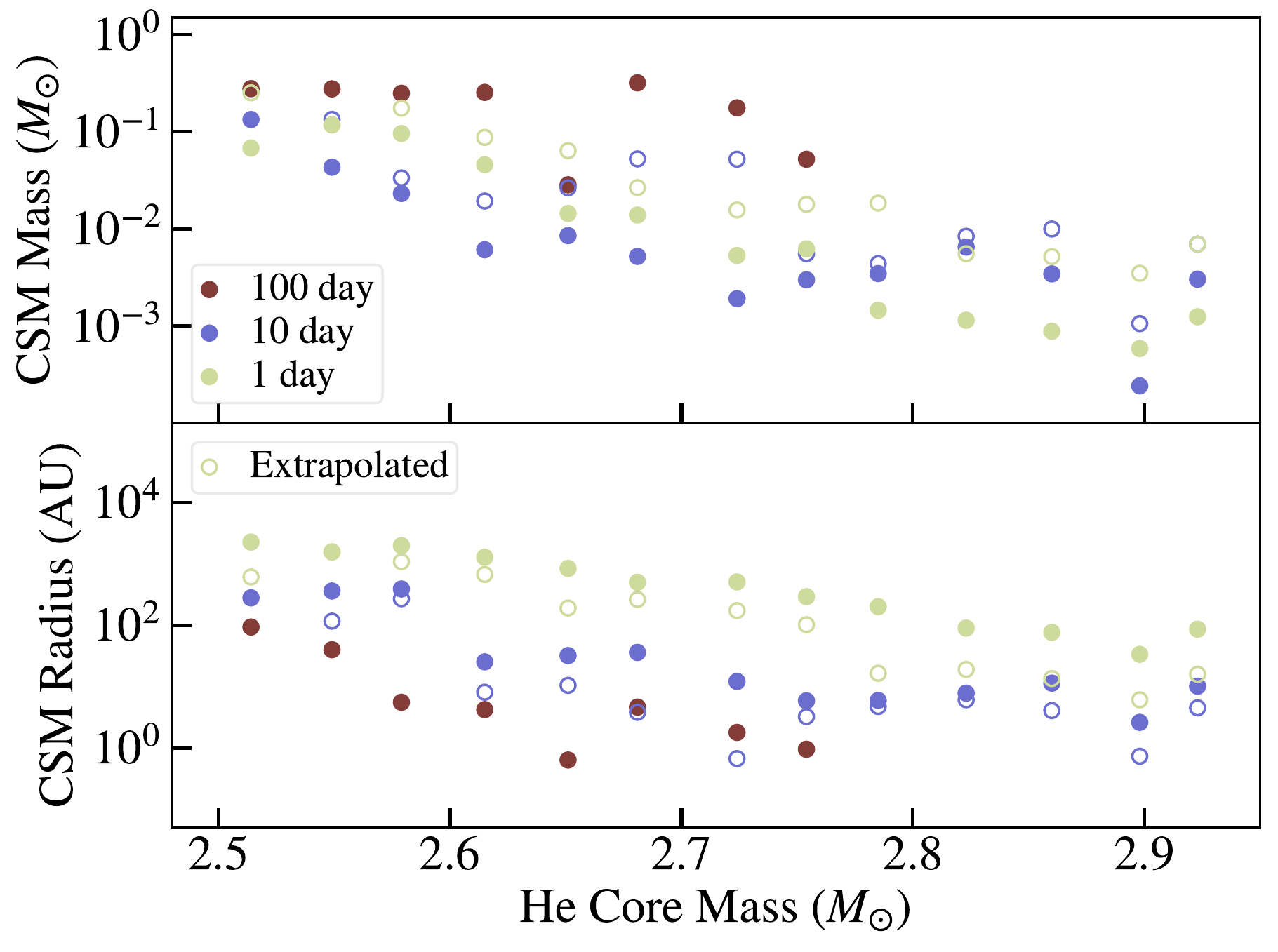}
    \caption{Properties of each binary system undergoing late-stage mass loss. Points are plotted as a function of the helium star's initial mass, and colors correspond to the initial orbital period given in the legend. Dots represent mass loss up to the end of the solid lines in Figure \ref{fig:mdots}, and open circles include the extrapolated mass loss shown as dotted lines in Figure \ref{fig:mdots}. Top left: Time before Si-burning when late-stage MT ensues,~$t_{\rm MT}$, as defined in Section \ref{sec:binaryevol}. Bottom left: Final mass of each stripped star after case BB and late-stage MT.  Right: Predicted mass (top) and radius (bottom) of CSM due to late-stage mass loss. 
    }
    \label{fig:csmprops}
    
\end{figure*}

\section{Results}

\subsection{Single Star Evolution}

Stripped stars with initial masses $2.5\, M_{\odot} \lesssim M_{\rm He} \lesssim 3\, M_{\odot}$ expand during several phases of their evolution. In the left panel of Figure \ref{fig:singlervst}, the radii of the stripped stars increase by two orders of magnitude during C-burning beginning $\sim \! 10^5$ years before Si-burning, which in a binary system causes case BB MT. Notably, the stars contract after C-shell burning (a few decades before Si-burning) and expand by a factor of a few again during O/Ne-burning, which can initiate late-stage MT.

The right panel of Figure \ref{fig:singlervst} plots the time before Si-burning when this late expansion occurs versus stripped star mass $M_{\rm He}$. For higher masses, expansion occurs months before Si-burning during late O-burning, but lower masses expand again a decade before Si-burning. However, the expansion is driven not by core burning but rather by intense He-burning at the base of the helium envelope. This behavior can be understood by the mirror principle \citep[e.g.,][]{Kippenhahn2012,Laplace2020}, in which core contraction after certain burning phases (e.g., C-shell burning) causes the temperature of the He-burning shell to increase. As the temperature-sensitive triple-alpha energy generation rate increases significantly, the envelope of the star expands in response to the intensified heating at its base. 

From the maximum radius to which each stripped star expands, we estimate the maximum orbital period~$P_{\rm orb}$ for the star to fill its Roche lobe during late-stage MT.
For a companion mass of~$M_{\rm c}$, the mass ratio is~$q = M_{\rm c}/M_{\rm He}$. Then the ratio of Roche lobe radius~$R_{\rm RL}$ to semi-major axis~$a$ is approximately \citep{eggleton1983}
\begin{equation}
\label{eq:roche}
    \frac{R_{\rm  RL}}{a} = 0.49 \frac{q^{2/3}}{0.6q^{2/3}+\log(1+q^{1/3})} .
\end{equation}
Setting each star's maximum radius during O/Ne-burning to $R_{\rm RL}$ and applying Kepler's third law gives the maximum orbital period for Roche lobe overflow,~$P_{\rm orb, max}$. In the right panel of Figure \ref{fig:singlervst}, points are shaded by the value of~$P_{\rm orb, max}$, which tends to decrease with He mass. The stripped stars may initiate late-stage MT up to orbital periods of months to years.


\subsection{Binary Evolution}
\label{sec:binaryevol}

Figure \ref{fig:mdots} shows the mass transfer rates~$\dot{M}$ and accumulated mass loss of our binary models at $P_{\rm orb}=100,\, 10,$ and $1$ day. Once the models achieve very high mass loss rates~$\dot{M} \gtrsim 10^{-3}\, M_{\odot}/\rm{yr} $, MESA systematically encounters numerical difficulties at the surface of the star, where mass layers are rapidly stripped. In models where MESA is unable to evolve the mass transfer up to Si-burning, we estimate the time until Si-burning by comparison with single-star models and also extrapolate further potential mass loss by assuming that~$\dot{M}$ plateaus until Si-burning.
This approximates the behavior of models that do evolve to Si-burning (e.g.~$P_{\rm orb} = 100$ day); in practice, the extrapolation may be a lower limit to the true mass loss since mass transfer rates are usually increasing sharply when models terminate.
Binary models at larger~$P_{\rm orb}$ begin late-stage MT later in the donor's lifetime, and the highest-mass models~$M_{\rm He} \gtrsim 2.8\, M_{\odot}$ do not expand enough to fill their Roche lobes at~$P_{\rm orb} = 100$ day. At $P_{\rm orb} = 1$ day, models do not fully detach from their Roche lobes after C-burning, but late-stage MT clearly manifests as $\dot{M}$ increases by $\sim$2-3 orders of magnitude during O/Ne-burning.

Typical mass loss rates are
$10^{-3}\text{--}10^{-1}\, M_{\odot}/\rm{yr}$ during late-stage MT. Though our binary models at~$P_{\rm orb} =100$ day rise to~$\dot{M} \sim 0.1\text{--}1\, M_{\odot}/\rm{yr}$ in the last weeks to months before Si-burning, these highly uncertain values occur because the models greatly overfill their Roche lobes during these phases, causing the mass transfer scheme in MESA to break down. 
We define the time until Si-burning when late-stage MT occurs,~$t_{\rm MT}$, when~$\dot{M} > 5\times 10^{-4}\, M_{\odot}/\rm{yr}$, significantly exceeding the case BB MT rate of~$\sim 10^{-4}\, M_{\odot}/\rm{yr}$.
In the top left panel of Figure \ref{fig:csmprops},~$t_{\rm MT}$ is shown for each He star.
The mass loss rate tends to rise months to years before Si-burning for models of larger mass and longer~$P_{\rm orb}$, but late-stage MT can occur years to decades before Si-burning for lower mass and shorter~$P_{\rm orb}$. 

The bottom left panel of Figure \ref{fig:csmprops} shows the final masses after both case BB and late-stage mass loss, which range between~$\sim \! 1.4\text{--}2.9\, M_{\odot}$. 
The low pre-collapse masses imply small SN ejecta masses~$\lesssim 1.5\, M_{\odot}$, assuming~$M_{\rm NS} = 1.4\, M_{\odot}$. 
Compared to similar models in \cite{Tauris2015}, our mass transfer rates during C burning and final masses are consistent with their results. Following their argument that models with final CO core masses $\gtrsim 1.43\, M_{\odot}$ will reach iron core collapse, we expect that our lowest-mass models $2.5\text{-}2.55\, M_{\odot}$ may become electron-capture SNe, while the majority of our models $\gtrsim \! 2.6\, M_{\odot}$ will undergo core collapse. The final fate of our models corresponds to slightly different initial He core masses than in \citet{Tauris2015}, as our stellar evolution implementation produces slightly higher CO core masses for the same initial mass.


\subsection{CSM Properties}
\label{sec:MdotCSM}
To estimate the properties of CSM ensuing from late-stage MT, we treat each donor's mass loss as ejected from the system in the vicinity of the accretor. Stable MT at these high rates may form an advection-dominated, geometrically thick accretion disk around the companion that can drive a large proportion of mass from the outer disk, lost through the L2 point \citep{lu2022,pejcha2016}. 
Motivated by this scenario, we assume lost mass leaves with the orbital velocity at the L2 point. In reality, the ejection speed may vary due to initial conditions and torquing by the binary, and ejection velocities and CSM radii smaller by a factor of~$\sim$3 may be more realistic \citep[][]{hubova2019}.

Shells of expelled material form at a distribution of radii around the system, so we perform a mass-weighted average of these radii to calculate the characteristic CSM radius. The integrated mass loss rate at core collapse equals the total CSM mass in each system. 
As shown in the right panel of Figure \ref{fig:csmprops}, we predict CSM masses ranging from~$10^{-3} \, M_\odot$ for~$\sim \! 2.9\, M_\odot$ progenitors up to~$\sim \! 3\times 10^{-1}\, M_{\odot}$ for~$\sim \! 2.5\text{--}2.7 \, M_\odot$ progenitors. 
As it originates from stripping of the He envelope, the CSM produced by our models is He-rich, with He mass fractions $\gtrsim 0.7$ for the majority of the CSM mass.

In our models, the orbital velocity at L2 increases from~$\sim \! 100$ km/s at~$P_{\rm orb} \! \approx \! 100$ day to~$\sim \! 500$ km/s at~$P_{\rm orb} \! \approx \! 1$ day. Mass ejected from the system at these velocities reaches radii of~$\sim 1\text{--}10^4$ AU. Lower mass and shorter $P_{\rm orb}$ models tend to produce CSM at larger radii, as late-stage MT begins earlier in the evolution and in the latter case is ejected with larger velocities.

\subsection{Common Envelope Events}
\label{sec:CEE}
\begin{figure}
    \includegraphics[width=\columnwidth]{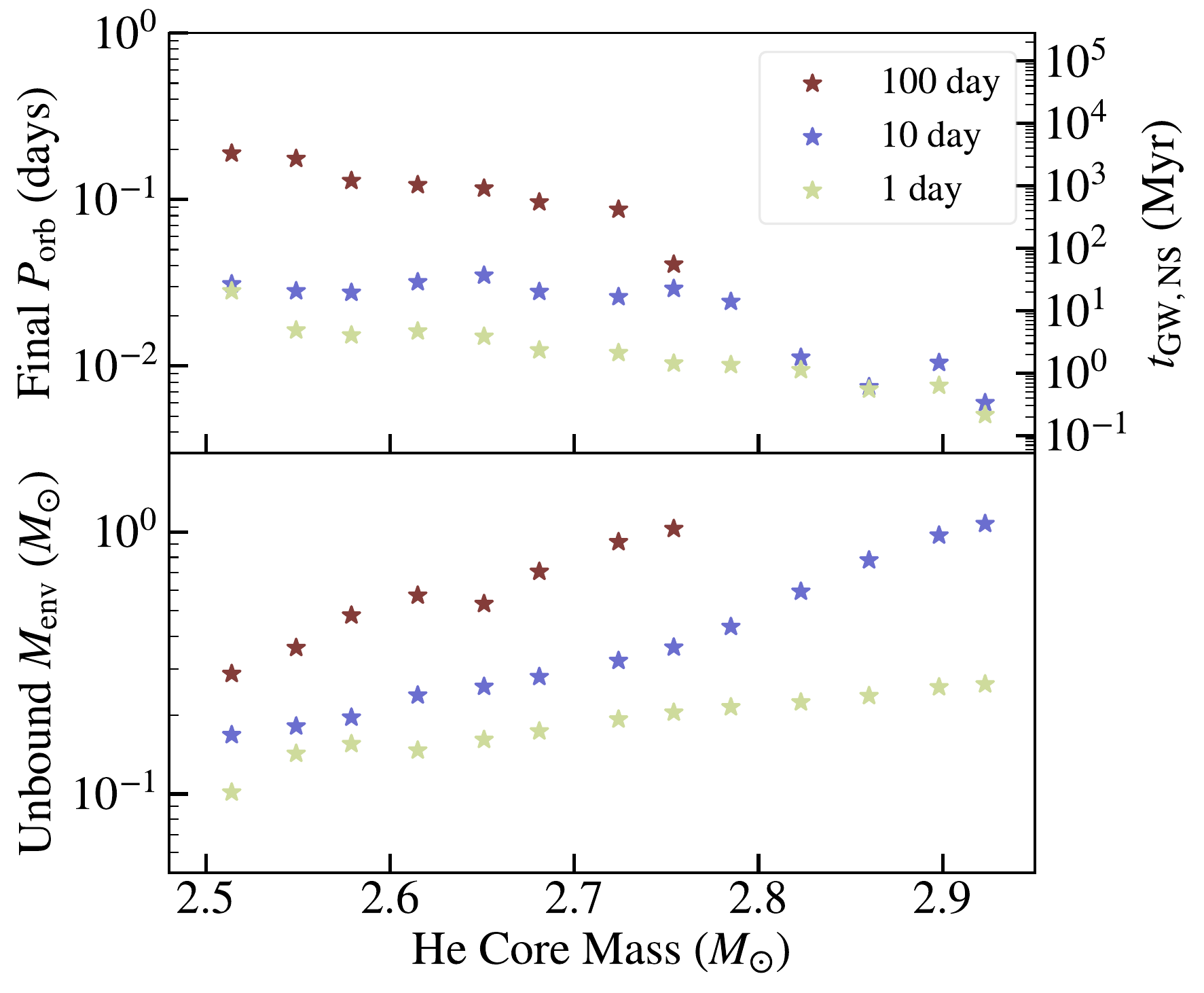}
    \caption{Properties of the binary systems after a common envelope event (CEE), assuming unstable MT begins once $\dot{M} > 5\times 10^{-4}\, M_{\odot}/\rm{yr}$ and that inspiral ends once the NS unbinds envelope down to the CO core. Points are plotted as in Figure \ref{fig:csmprops} as a function of each stripped star's initial He core mass. Top: Values on the left axis show the final orbital period of each system after CEE. The right axis values are the gravitational wave merger timescales for a binary system consisting of two neutron stars orbiting at the periods of the left axis. Bottom: Envelope mass unbound by the CEE.
    }
    \label{fig:ceprops}
    
\end{figure}

The sharply rising mass transfer rates in all our models may indicate the onset of unstable MT, leading to a common envelope event (CEE). In this case,
we would expect the companion to inspiral into the envelope of the ultra-stripped star, with total mass $M_{\rm s}$ at the onset of CEE. We predict the outcome by assuming that unstable MT ensues soon after~$\dot{M}$ exceeds $5\times 10^{-4}\, M_{\odot}/\rm{yr}$, and that the inspiral will terminate once the change in the orbital energy is sufficient to unbind the entire envelope of mass $M_{{\rm s}, \rm env}$ exterior to the C/O core of mass $M_{{\rm s},\rm core}$. To quantify this, we use the $\alpha$ energy formalism:

\begin{align}
    E_{\rm bind} & = \alpha \Delta E_{\rm orb} \\ 
    & = \alpha\left( -\frac{G M_{\rm s} M_{\rm c}}{2 a_{\rm i}} +\frac{G M_{{\rm s},\rm core} M_{\rm c}}{2 a_{\rm f}} \right) 
\end{align}
where~$E_{\rm bind} = \int_{\rm core}^{\rm surface} -\frac{G m}{r} + \epsilon(m)\, dm$, where $\epsilon$ is the specific internal energy. Here, the CE efficiency $\alpha$ parameterizes the fraction of orbital energy used to eject the envelope, $M_{\rm c}$ is the companion mass, 
and $a_{\rm i}$ is the initial orbital separation, determined by Equation \ref{eq:roche} with $R_{\rm{RL}}$ equal to the stellar radius at CE onset.
We solve for the final orbital separation $a_{\rm f}$ which satisfies this equation for each binary model assuming $\alpha = 0.3$, consistent with observational constraints \citep{zorotovic2010,zorotovic2022}. Though defining the final mass after the CEE is uncertain, we find that our results are not sensitive to this choice; here, we set the final mass to be the CO core mass at the onset of CEE.

Figure \ref{fig:ceprops} shows final orbital periods of our models after the CEE. In the vast majority of systems, the binary exits the CEE at orbital periods $\lesssim\! 4$ hours, which can merge within a Hubble time.  This orbital separation is too small to admit a main sequence star, so unstable mass transfer with a main sequence companion will likely result in a stellar merger followed by an unusual supernova. White dwarf or neutron star companions, however, can likely eject the envelope before merging with the CO core. The gravitational wave orbital decay timescale for a binary of two neutron stars $M_{\rm NS} = 1.4\, M_{\odot}$ at each orbital period, $t_{\rm GW, NS}$, is shown for comparison on the right axis. Models with larger $M_{\rm He}$ and shorter initial $P_{\rm orb}$ can reach final $P_{\rm orb}$ of under thirty minutes, corresponding to $t_{\rm GW, NS} \lesssim\! 10$ Myr. 

The bottom panel of Figure \ref{fig:ceprops} shows the mass of the unbound envelope due to the CEE, which increases with $M_{\rm He}$ from $\sim \! 10^{-1}\, M_{\odot}$ to $1\, M_{\odot}$ for more massive progenitors. These CSM masses typically exceed our estimates for the stable mass transfer scenario by a factor of $\sim$10, though we reiterate that those values are likely to be lower limits in many cases. We estimate the CSM radii produced by a CEE by assuming the envelope is ejected with a terminal velocity equal to the star's pre-CE surface escape velocity (consistent with $\alpha \sim 1/3$).

\begin{figure}
    \includegraphics[width=\columnwidth]{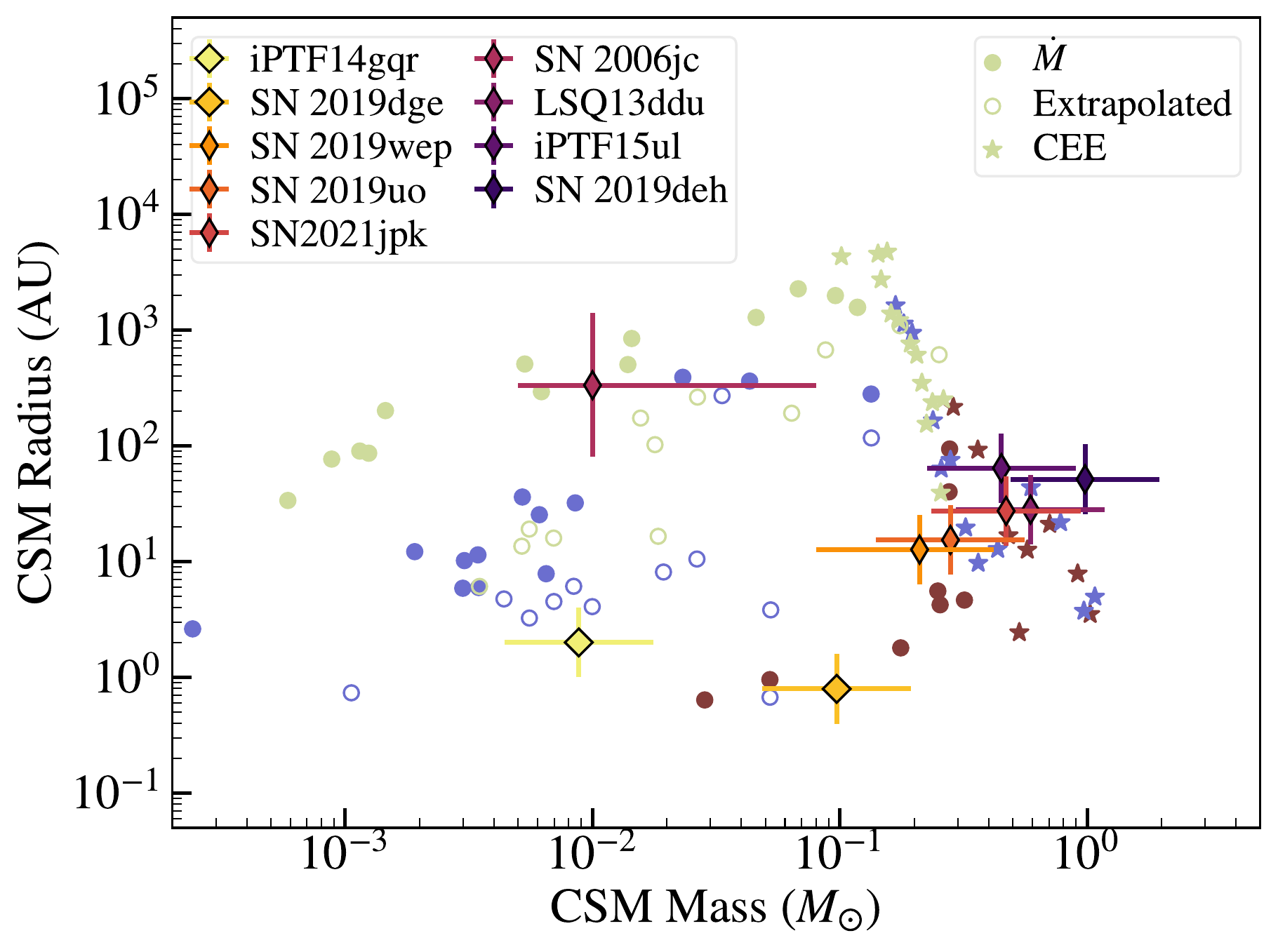}
    \caption{Dots and open circles are predicted mass versus radius of CSM due to late-stage mass loss (Figure \ref{fig:csmprops}, Section \ref{sec:MdotCSM}). Star symbols are predicted mass and radius of material unbound during CEE (Figure \ref{fig:ceprops}, Section \ref{sec:CEE}). Colors indicate initial~$P_{\rm orb}$ as in Figure \ref{fig:csmprops}. Points with error bars are estimated CSM properties of USSNe (square diamonds) and type Ibn SNe (thin diamonds).
    All are given error bars of at least a factor of 2 in each direction to account for systematic uncertainties in the modeling. 
    }
    \label{fig:csmpropsobs}
    
\end{figure}

\subsection{Comparisons to USSNe and type Ibn/Icn SNe}

Figure \ref{fig:csmpropsobs} compares CSM masses and radii inferred for several interacting SNe with our model predictions. We include our estimates for stable late-stage mass loss (Section \ref{sec:MdotCSM}) and from mass ejected due to a CEE (Section \ref{sec:CEE}). In general, methods of observationally constraining the mass and radius of CSM are likely uncertain by a factor of a few \citep{De2018,chatzopoulos2013,chatzopoulos2012}, so we show inferred values with error bars of at least a factor of 2 in each direction.

In the vast majority of our models, the predicted CSM is H-poor and He-rich, so interaction with the CSM during a SN would likely produce spectra classified as type Ibn. Several type Ibn SNe are shown in Figure \ref{fig:csmpropsobs} as thin diamonds. SN 2006jc \citep{anupama2009} may be matched by a range of $P_{\rm orb}=1$ and $10$ day models for late-stage MT, whereas CSM estimates for iPTF15ul, SN 2019wep, and SN 2019uo \citep{pellegrino2022} fall within the late-stage MT estimates for initial~$P_{\rm orb} = 100$ day. All type Ibn events shown, including SN 2019deh, SN 2021jpk, and LSQ13ddu \citep[][]{pellegrino2022,clark2020}, 
can be explained by CSM produced in a CEE in several $P_{\rm orb} = 1,\, 10$, and $100$ day models.

Estimates from modeling shock-cooling emission of extended material around the progenitors in USSNe iPTF14gqr and SN 2019dge, shown as the blue and and orange diamonds, are consistent only with our smallest CSM radii predictions. Our models for late-stage MT with initial~$P_{\rm orb} = 10$ day and~$M_{\rm He}= 2.6\text{--}2.9\, M_{\odot}$ can explain the envelope mass~$M_e\approx 0.01\, M_{\odot}$ and radius~$R_e \gtrsim 2\, \rm{AU}$ derived for iPTF 14gqr \citep{De2018}. Similar-mass
late-stage MT models with initial~$P_{\rm orb}= 100$ day can explain observed material at~$\sim \! 1\, \rm{AU}$ for SN 2019dge \citep[][]{Yao2020}.

In addition, outer CSM regions located beyond a few to tens of AU have been detected in iPTF14gqr and SN 2019dge. Estimates from He II line emission of the mass of helium in this outer CSM provide lower limits of $\gtrsim\, 3\times 10^{-5}\, M_{\odot}$ for SN 2019dge and $\gtrsim\, 10^{-2}\, M_{\odot}$ for iPTF14gqr. CSM produced from late-stage mass transfer fits well with the properties of outer regions of CSM in both USSNe. In addition, our models routinely attain the inferred pre-supernova mass loss rates of ~$\gtrsim 10^{-2}\, M_{\odot}/\rm{yr}$ and~$\gtrsim 10^{-4}\, M_{\odot}/\rm{yr}$ respectively for iPTF14gqr and SN 2019dge.

Interaction with CSM has been detected in several type Icn SNe \citep[][]{pellegrino2022b,galyam2022}. These events mainly show narrow C/O emission lines, though none are conclusively devoid of He, and the type Icn SN 2019jc in particular has an He II feature. With the exception of the values derived for SN 2019hgp by \citealt{galyam2022}, the type Icn SNe tend to produce more massive CSM than our models predict. 
Most importantly, these events likely require lower He mass fractions than ejected by our models \citep{dessart2022}, which typically have $X_{\rm He} \! \sim \! 0.8$ and $X_{\rm CO} \! \sim \! 0.2$.



\section{Discussion and Conclusions}
\label{sec:discussion}

The fiducial scenario addressed by our binary models describes a He star, formed from the initially less massive star (the secondary) in a binary, with a NS companion evolved from the initially more massive star (the primary). However, primary stars within our modeled mass range will exhibit the same behavior. If the primary has a low-mass main sequence (MS) companion, case B MT is expected to be dynamically unstable leading to CEE. The low-mass MS star could survive the inspiral and exit CEE in a close orbit with the He star -- these are likely the progenitors of low-mass X-ray binaries \citep[e.g.,][]{1993ARA&A..31...93V, kalogera1998}. If the companion is massive, case B MT is likely stable and may widen the orbit, but for post-MT separations less than a few $100\, R_{\odot}$ the He star can still overfill its Roche lobe during late-stage expansion. Thus the late-stage mass transfer displayed by our stripped star models may affect the appearance of a Type Ib/c SN coming from either primary or secondary stars with $M_{\rm He}\sim 2.5\text{--}3\, M_{\odot}$.

At the extreme mass transfer rates predicted, the dynamics of the ejected mass are uncertain. Since the donor may greatly overfill its Roche lobe, mass may also flow out of the donor's outer Lagrange point (L3 if the donor is more massive; \citealt{linial2017,marchant2021}).
Even if the companion is not a compact object, the high mass transfer rates, if stable, may form a geometrically thick accretion disk around the companion. The disk will be super-Eddington even at large radii, such that L2 mass loss is predicted \citep{lu2022}.
The ensuing circumbinary outflow may cause appreciable additional angular momentum loss given the larger lever arm of the L2 point. This effect may shrink the orbital separation more rapidly, increasing and potentially destabilizing the mass transfer rates. In preliminary tests, we have noticed a $\sim \! 20\%$ increase in the MT rate if we change the specific angular momentum of the mass lost to that of the L2 point.

However, the ejection velocities of $\sim\! 200$ km/s predicted in this framework tend to be lower than estimated from line widths in observed SNe. In the case of accretion onto a compact object, the disk around the accretor may launch a super-Eddington wind that sweeps up the slower outflow from the L2 point. It is also unclear how the ejected mass will be distributed. Though we report only a single CSM radius, the material will certainly cover a large radial extent and may not have a smooth or spherically symmetric profile.

If the mass transfer becomes dynamically unstable, the accretion disk scenario is superseded by a common-envelope event, as explored in Section \ref{sec:CEE}. Our models with~$P_{\rm orb} = 100$ day appear highly susceptible to CEE, as they reach very high mass transfer rates $\dot{M}\sim 1\, M_{\odot}/\rm{yr}$ which approach the dynamical regime of mass transfer. They also have fully convective envelopes at the onset of late stage MT and therefore are more inclined to lose MT stability. Once~$P_{\rm orb}$ decreases to $1$ day, models instead host only a very thin surface convective region. Hence, late-stage MT in binaries at long $P_{\rm orb}$ may result in CEE events that eject $\sim \! 1 \, M_\odot$, while binaries at shorter $P_{\rm orb}$ may remain stable due to their mostly radiative envelopes. The former may account for many of the observed type Ibn SNe, while the latter may account for USSNe with He-rich CSM. Both scenarios may contribute to NS mergers, depending on the degree of orbital decay during CEE or late stage MT.

We roughly estimate the birth rate of progenitor systems that exhibit late-stage MT in order to compare with the rate of type Ibn SNe that they may produce.
The volumetric rate of type Ib/c SNe is $\sim\! 2.5\times 10^{-5}\, \rm{Mpc}^{-3}\, \rm{yr}^{-1}$ \citep{li2011,frohmaier2021}.
Given that type Ib/c SNe are thought to arise from binaries with a stripped star component, the type Ib/c rate approximates the birth rate of such systems, regardless of whether the primary or secondary star produces the SN.
Our He star models could produce late-stage MT as either the primary or secondary star, and they represent the low-mass subset of type Ib/c SN progenitors. 
Binaries producing late-stage MT include at least one star with $M_{\rm ZAMS} \sim 13\text{--}15\, M_{\odot}$, whereas we assume systems contributing to the ordinary type Ib/c rate contain at least one star with $M_{\rm ZAMS}\gtrsim 15\,  M_{\odot}$.  By integrating the IMF \citep{kroupa2001}, we find that systems with late-stage MT constitute~$\sim \! 11\%$ of type Ib/c SN progenitors. Thus we estimate a birth rate for systems that exhibit late-stage MT of~$\sim 3\times 10^{-6}\, \rm{Mpc}^{-3}\, \rm{yr}^{-1}$.
To compare to the rates of type Ibn SNe, we note that the ZTF catalog estimates~$\sim$10 type Ibn SNe per~$\sim$900 CCSNe \citep{perley2020}. \citet{maeda2022} estimate $\sim \! 1$\% of CCSNe are type Ibn SNe, giving a volumetric rate of $\sim \! 10^{-6}\, \rm{Mpc}^{-3}$, though these rates may be underestimates since such brief transients can be missed by surveys. Moreover, the calculation above likely overestimates the birth rate, as some massive star binaries may evolve to wide separations where late-stage mass transfer does not occur, and some type Ib/c SNe may originate from merging systems. Thus the birth rate of our progenitor systems appears to be roughly compatible with the type Ibn SN rate.

At the high mass transfer rates seen in all our models, there is a very high degree of Roche lobe overflow that the MESA mass transfer schemes do not capture well. More detailed modeling will be necessary to quantify more accurate late-stage mass transfer rates. Nevertheless, the values presented here are conservative estimates for models where we see the mass transfer rates increasing towards the end of our simulations. Additional sources of angular momentum loss not modeled here will serve only to exacerbate Roche lobe overflow through faster orbital decay. Ultimately, late-stage mass transfer initiated during O/Ne-burning will unavoidably lead to extremely high mass transfer rates that can considerably influence the properties of these binary systems in the final years before core collapse.

\section*{Acknowledgments}

We thank Thomas Tauris for helpful discussion, and Pablo Marchant for guidance in implementing the corrected mass transfer prescription. This material is based upon work supported by the National Science Foundation Graduate Research Fellowship under Grant No. DGE‐1745301.

\bibliography{bib}
\end{document}